# Not Quite 'Ask a Librarian': AI on the Nature, Value, and Future of LIS

**Jesse David Dinneen**
Humboldt-Universität zu Berlin, Germany
*jesse.dinneen@hu-berlin.de*

**Helen Bubinger**
Humboldt-Universiät zu Berlin, Germany
*helen.bubinger@student.hu-berlin.de*

## ABSTRACT

AI language models trained on Web data generate prose that reflects human knowledge and public sentiments, but can also contain novel insights and predictions. We asked the world's best language model, GPT-3, fifteen difficult questions about the nature, value, and future of library and information science (LIS), topics that receive perennial attention from LIS scholars. We present highlights from its 45 different responses, which range from platitudes and caricatures to interesting perspectives and worrisome visions of the future, thus providing an LIS-tailored demonstration of the current performance of AI language models. We also reflect on the viability of using AI to forecast or generate research ideas in this way today. Finally, we have shared the full response log online for readers to consider and evaluate for themselves.

## KEYWORDS

library and information science; artificial intelligence; foundations of information science; research methods

## INTRODUCTION

Some questions about the library and information science (LIS) community persist across decades, re-appearing perennially. Especially popular are questions about LIS's nature, identity, and place among other fields – and thus also the most suitable name for it – and the value it offers to society via education, knowledge production, and public services (e.g., Kaden *et al.*, 2021; Nolin & Åström, 2010). Similarly, researchers and practitioners are generally concerned with the future of LIS, asking for example what we should prepare for next as the discipline grows (Weller & Haider, 2007) or as rapidly changing technologies like artificial intelligence are introduced into our institutions (Fernandez, 2016). While a descriptive answer can be given with statistics (e.g., from associations' databases tracking who works where and on what topics), prescriptive and speculative answers can be (and have been) provided by, for example, reflective editorials, persuasive papers, and panel sessions soliciting the forecasting of experts (e.g., sessions at ASIST 2019 on LIS identity and the need for foundations in LIS).

An additional approach has recently become possible, which integrates description and speculation: asking artificial intelligence what LIS is and could be. The approach is a hybrid insofar as AI trained on public data can reflect the status quo of human knowledge on a topic (i.e., descriptive), but may also process that data in a way that produces novel and interesting ideas, for example by further developing existing perspectives or implicitly combining perspectives to surprising effect. Notably, the AI system GPT-3 has produced novel and sophisticated commentary on philosophers' writings about whether and what sense it could be said to be thinking or conscious, which included (apparent) self-reflection, and that commentary led to interesting further fruitful discussion among top philosophers (Weinberg, 2020). In other words, though AI-generated prose may be banal or useless, it may also be very interesting, and can thus consulting AI language models comprises a new method for generating commentary or forecasting as well as producing research ideas more generally (i.e., beyond the nature of LIS).

Although many creative uses of AI are acknowledged today (Anantrasirichai & Bull, 2021) and some worry has been raised about AI generating *fake* research (Dehouche, 2021), to our knowledge no prior work has assessed the ability of GPT-3, or any other AI language model, to generate *genuinely useful research ideas or commentary about a field*. We used Philosopher AI (https://philosopherai.com) to ask OpenAI's world-class language model, GPT-3, difficult questions about the nature, value, and future of LIS. The AI-generated responses, which we have shared online and the highlights of which we discuss below, tell us about the written public record about LIS, provide novel perspectives on perennial issues that can fuel further discussion at the ASIS&T annual meeting, and provide the LIS community with a tailor-made (and sometimes entertaining) demonstration of the current state and limitations of AI language models (i.e., it demonstrates the quality of the prose of today's best AI). After discussing the responses we also reflect on the all the responses and the experience of generating and reviewing them to consider what we perceive to be the usefulness and practical feasibility of using AI for such purposes today, constituting a preliminary evaluation of a new and increasingly viable method.

## METHOD

Generative Pre-trained Transformer 3, or GPT-3, is a language model developed by OpenAI in 2020 (now licensed exclusively by Microsoft) that uses deep learning to identify features of inputted text, modelled as 175 billion parameters in a neural network, and when prompted it generates *new* text with similar features and distributions (of occurrences of phrases, ideas, synonyms, and so on) as seen in its training data. GPT-3 was trained with data that can be characterised as human-generated prose, code, mathematical formulae, and so on from various Web sources (e.g., WebCrawl and Wikipedia; Weinberg, 2020). For its data and complexity GPT-3 is considered world-class; it can generate convincing fake news, entertaining fiction, poetry, do mathematical analysis, and write code, all of which reflect its training data – and thus published human knowledge and opinion – but are also often novel, interesting, entertaining, and so on (Dickson, 2020; Diresta, 2020). It has thus attracted considerable attention from press and academe. For a longer introduction to GPT-3's origin, capabilities, and societal and philosophical implications, see Floridi and Chiriatti (2020).

At the time of writing, public access to GPT-3 is provided through different platforms whose developers have permission to use its API (often for a fee). We used Philosopher AI (pictured in Figure 1), an open-source tool with a simple Web interface that allows the user to input a query (i.e., prompt) to GPT-3 and get a detailed text response. As developers pay to access the GPT-3 API, sites/tools like Philosopher AI generally charge users a subscription or, in the case of Philosopher AI, per query (at about $3.33 US each), and some implement their own features (e.g., query processing and response filtering).

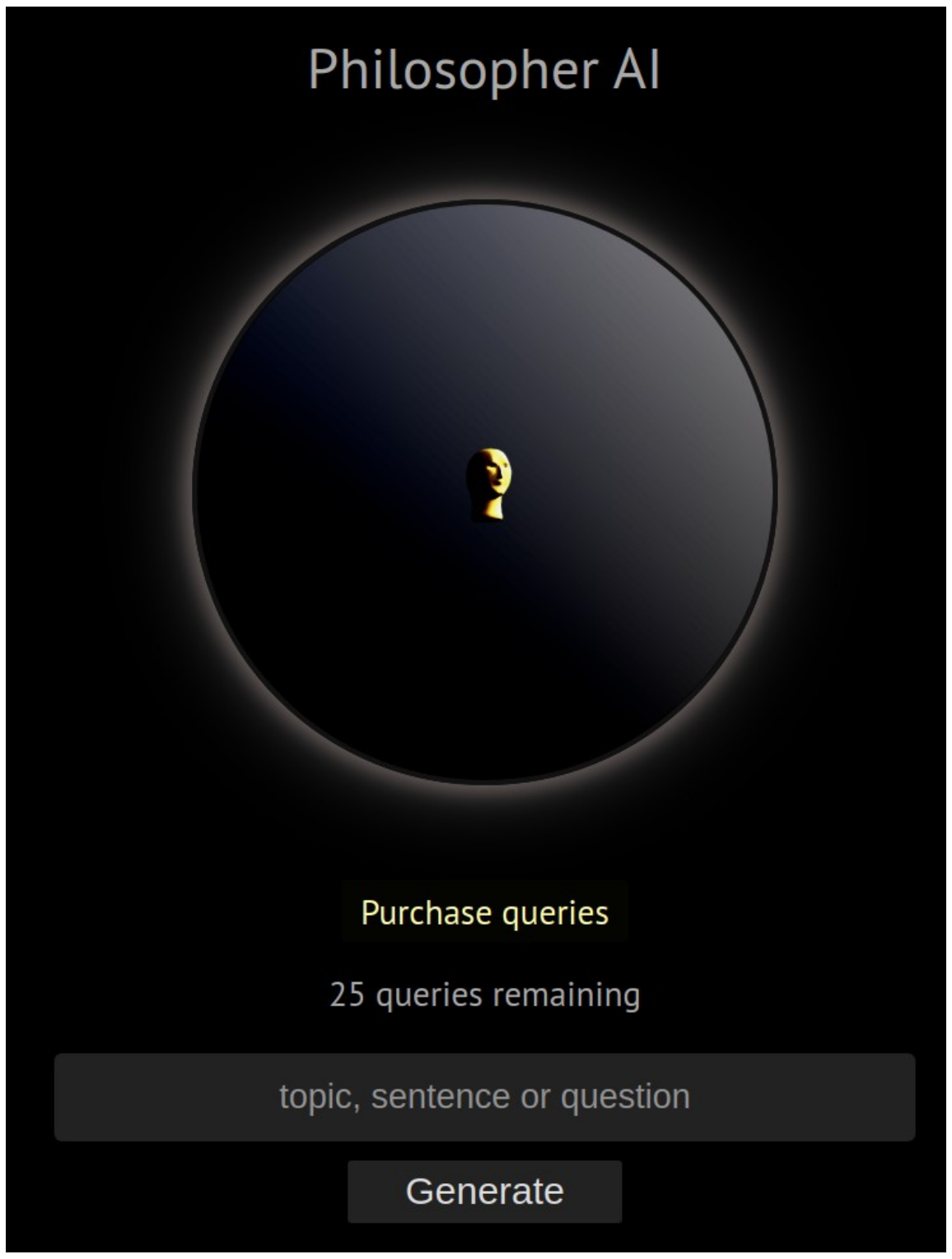

**Figure 1. The Philosopher AI interface to GPT-3**

To generate questions to pose Philosopher AI, we consulted immediate colleagues in LIS, reviewed seminal literature (cited above and in the discussion), and extracted topics described as 'big questions' or 'grand challenges' in conversations at recent international LIS meetings and venues, for example the 2021 iConference panel on the iSchools' identity (Kaden *et al*., 2021) and the 2020 ASIST EU Chapter *Uncommons Session. S*imilar questions were merged and rephrased with common general terminology (e.g., 'what exactly is LIS' and 'what is LIS like' are replaced by 'what is the nature of LIS'). The result is fifteen questions, used verbatim as prompts, that together address LIS's: **nature** (and thus the best name for it), **value**, and **future**, with the future including a particular focus on the role of AI. Table 1 presents the prompts, grouped by topic, together with the number of queries required to get three usable answers.

| Topic | Prompts | # queries to get 3 usable answers |
|---|---|---|
| Nature of LIS | 1. what is the nature of 'library and information science'? | 4 |
| | 2. what kind of science is 'library and information science'? | 6 |
| | 3. where does 'library and information science' fit among the academic disciplines like humanities, social sciences, natural sciences, and so on? | 6 |
| | 4. what makes 'library and information science' unique as a field of study? | 7 |
| | 5. which subfields are at the core of the discipline 'library and information science', and which are at the periphery? | 5 |
| | 6. is 'library and information science' the best name for that field? | 5 |
| | 7. what is the best label or name for the field that studies information and information institutions? | 5 |
| | 8. what is the best label or name for the field that studies the intersection of information, people, and technology? | 4 |
| Value of LIS | 9. what is the societal value of the field of study known as 'library and information science'? | 3 |
| | 10. what does a degree in 'library and information science' prepare students to do? | 3 |
| Future of LIS | 11. what are the grand challenges that should concern the discipline 'library and information science'? | 4 |
| | 12. what are the biggest challenges facing the information society today? | 4 |
| | 13. what will libraries look like in 50 years? | 3 |
| | 14. how will artificial intelligence impact 'library and information science'? | 6 |
| | 15. how will artificial intelligence impact libraries? | 4 |

**Table 1. Specific prompts, posed to Philosopher AI, organised by topic**

To avoid encouraging a particular opinion in the AI's response we did not iteratively revise the prompts nor cherrypick from the results: after an initial test to establish if the terms would encourage on-topic answers, we put in queries and discarded only responses that were not usable for meaningfully commenting on the question (i.e., neither answered the question nor discussed anything related in a coherent way). Most questions required only four queries to produce three usable answers (mean 4.6, max. 7), as discussed below. No responses were rejected for their direction (i.e., positive or negative opinion expressed). While our presentation of the results is necessarily influenced by our individual backgrounds and interests, we tried to minimise the effect of this by first independently generating our impressions of the responses, and then checking them for overlap (i.e., inter-subjective agreement); overlap between researchers was very high, with most summaries of the responses being nearly identical and detailed impressions being similar. To increase transparency of our analysis we provide quotes with reference to the numbered prompts in the full response log, shared at https://github.com/jddinneen/ai-results.

## RESULTS & DISCUSSION

Here we summarise Philosopher AI's responses to our questions, grouped by topic (nature, value, and future of LIS). Quotes in this section are provided with citations referring to the numbered responses shared online (i.e., 2.1.2 refers to the second topic, first query, second response). At the end of the section we discuss trends across the topics and briefly evaluate the approach of seeking insight in AI today.

### The Nature of LIS

Questions about the nature of LIS typically required Philosopher AI 5 or more attempts (i.e., queries) to produce three usable responses.

Philosopher AI's perspectives on the nature of LIS focus heavily on libraries in particular. In its first response it acknowledged the topic is commonly observed to be "one of the most difficult topics to discuss", and that it ultimately "depends on who you ask" as "there is no one universal definition of what library and information science is" (1.1.1), but confusingly, also noted that librarians and non-librarians agree that LIS is the "study of libraries and all related activities". It also suggested the answer may lie in understanding what libraries are, how they might be categorised, and how one defines information, reminiscent of the task of defining digital libraries (Borgman, 1999). To wit, it suggests a library is "a place of knowledge" that "contains information that can be useful to the patrons who come in for various reasons" but which has "changed from being a source of knowledge to some place where people come in to read or use their phones" (1.1.3), and categorises libraries into five categories: public, private, of different sizes, research, and confusingly, museums. The definition of information provided was: public or private facts or data (1.1.2). Though consistent with existing definitions of information in LIS (Dinneen & Brauner, 2015), the perspective was too short on detail to discuss further.

When asked more specifically about *what kind of science* LIS is, Philosopher AI produced as many unusable responses as usable ones. One response will likely sound familiar to information scholars: "a branch of science that focuses on the collection and organisation of knowledge" or more specifically "a kind of social science focusing on the collection, organisation, classification, preservation and dissemination of recorded human knowledge", and "the study of human behaviour in relation to information" (1.2.1). It added "the main goal of this field is to ensure the preservation and dissemination of recorded human knowledge for future generations. I believe that LIS is a scientific discipline with its own scholarly journals and annual conferences where people from all over the world get together to exchange views on various topics related to this field" (1.2.1). True enough, but not especially novel nor inclusive of the full nuance, complexity, and variety of the field (c.f. Bates 1999; Buckland, 1999). In another answer, it combined aspects of knowledge organisation and data management in an unexpected way: "library and information science is about the management of data", which it argues is done through classification systems (with reference to Dewey and MeSH): "Without these systems, there would be no way to organize the vast amounts of data that are on servers all over the world" (1.2.3). Under some interpretations this may be true, and it perhaps broadly aligns with Otlet's vision of LIS as a discipline classifying the world of facts (Rayward 1994), but may also sound counterintuitive to readers accustomed to classifying *works or items* rather than the data that represent them or their surrogate records. The least focused answer considered the purpose of libraries, what wisdom aliens may have, and if anybody really understands how Websites work (1.2.2), which we take to say more about GPT-3 than LIS. Notably missing from the answers is an acknowledgement of how the nature of LIS changes over time and with ever-changing turns, paradigms, etc (Hartel, 2019) and any mention of how theoretical commitments or professional values may distinguish it (Floridi, 2002; Foster & McMenemy, 2012).

Philosopher AI's answers regarding how or where LIS fits among other academic disciplines will likely match most readers' views. One response positioned LIS between the social and natural sciences, though closer to the former, e.g., characterising LIS as "a discipline that studies how people use the products of the natural sciences, namely knowledge", and "a sort of hybrid discipline between social sciences like economics or political science, on the one hand, and natural sciences like physics or chemistry" (1.3.2), echoing varied perspectives from the history of LIS (Buckland, 2012). Another answer put LIS "somewhere between the social sciences and humanities. It draws on both but seems to have a strong leaning towards the humanities", adding that LIS is a relatively new but "reasonably well-defined subject" and less scientific than other social sciences but with its own distinct features (1.3.3). While these perspectives are plausible, they were presented without rationale or conviction: "I think it's a distinct discipline, but I can see arguments on both sides" (1.3.3). Finally, in critiquing the name *LIS* (a topic which we return to below), the presence of a response was used to suggest that "as you can see, the AI has no prejudice towards any one discipline and is able to come up with conclusions that are not biased by human experience" (1.3.1). Considering most scholarly study points to the contrary (c.f. Ntoutsi *et al.*, 2020), and considering that public AI literacy is currently relatively low (Markazi & Walters, 2021), the appearance of such claims in the output of AI is worrisome.

Explaining what might make LIS a unique field of study required the most attempts (seven), to produce three usable responses, but none included direct, explicit answers, suggesting this was the most difficult question. In two answers (1.4.1, 1.4.3) it noted rather that *librarians* and information *scientists* are unique. For example, librarians play unique roles: as an intermediary, filter, or curator for/between the "library users" and information (1.4.1). To our knowledge it is a novel approach to explain the identity of LIS through exclusive reference to the relevant professional and research roles (i.e., suggesting librarians are what makes LIS unique); given the prior critique of the LIS name, perhaps Philosopher AI would argue the L in LIS is useful for making our field's unique identity apparent to those unfamiliar with the nuances of information science. We return to that debate below. Finally, it offered a distinction between *information science* (sans L) and computer science: the former focuses on helping people find information, while the latter focuses on creating that tools that are needed for the former (1.4.2). This is a reasonable distinction, but as an account of the field it does not sufficiently capture the variety of topics and focuses in LIS (e.g., while information retrieval and HCI fit nicely into the view, the most characteristic aspects of topics like personal archives, indigenous information behaviour, or knowledge organisation are not accounted for, just to name a few).

When asked about which subfields of LIS may be at the core rather than the periphery, one answer was explicit about the distinction: metadata, cataloguing, classification, data curation and preservation are at the core of LIS, whereas the periphery encompasses "everything else, such as rare books or digital libraries" (1.5.3). The other two answers did not distinguish the two. Another answer only identified and described two subfields of *librarianship*, collection development and reference services (1.5.2), whereas the last answer more generally described concerns of LIS, such as "the storage, retrieval, preservation, dissemination and organization of information" and even "all forms of communication" and "oral traditions such as storytelling" (1.5.1). It is perhaps unsurprising that no definitive answer was given, as the question is challenging even for LIS scholars (Bates, 2007).

*The best name for LIS* – Regarding the field's current name "Library and Information Science", two responses were critical. One argued the name is misleading, as is "information science", because *library* is not sufficiently broad to capture the "very diverse" field, whereas *information* is inaccurate because "it's not really about information at all but rather collection, organization, presentation and use of very diverse kinds of knowledge" (1.6.1). A more exhaustive answer said LIS "is a very bad name for the field" because LIS actually studies materials that hold knowledge or data (i.e., not information), "library" does not capture the many kinds of information storage places, "information" is neither specific nor unique enough to be helpful, and information science is closer to art than science (1.6.3). Though not entirely novel (c.f. Furner, 2015), the points each have merit and countering them requires a fairly sophisticated account of our field and what makes a good field name. A final answer avoided the L in LIS completely and stated that the name "information science" is a "fine" name, which suitably encompasses the wide variety of the many types of people in the field, and is unlikely to be confused with other fields, though it had some concern about the suitably and implied objectivity of the term "science" in a field comprised of many perspectives (1.6.2). Perhaps these points support the name *information studies*, which allows (but does not commit exclusively to) science, and does not favour one kind of information institution.

When asked about the best label for the field that studies *information and information institutions,* Philosopher AI took three distinct approaches in its answers. One was to emphasise and even exaggerate the *knowledge* aspect of the field, arguing for the name 'the field of knowledge', which it supposes contains the liberal arts, humanities, political science, psychology, and law, and which it confusingly states is both a broader field and a *subfield* of information science (1.7.1). While epistemologists may take exception to the suggested name, the perspective does reflect the nature of LIS as a meta-discipline (Bates, 1999, 2007). A second approach stated the best name for the field of all "libraries, archives, museums and other archival repositories of knowledge" simply *is* "librarianship", but there was little relevant support for the statement (1.7.2). The final approach avoided a direct answer but emphasised the importance of studying information itself, which it defined variously, because of its importance and many forms today (1.7.3). Perhaps the implicit proposal is to simply call the field (and perhaps our departments) 'information' in the same way other fields have done (e.g., history, philosophy, english, education).

When asked instead for the best label for the field that studies *the intersection of information, people, and technology* (a slogan used by several iSchools, for example on their Websites and in promotional materials), twice it instead critiqued the task itself. In one such case it simply discussed the difficulty of defining the term "information technologies" (1.8.3), whereas in the other it stated "one might as well ask what the name of physics should be, or mathematics, or even the whole of reality itself. *It's really just a way to avoid thinking about something more important by instead focusing on semantics*" (1.8.2). Some readers may sympathise with the commentary these anti-answers provide on the task and broader topic. The more straightforward answer was that "the best label is information studies or knowledge engineering" (1.8.1), but there too the AI was uncertain, adding that it was not sure it had anything interesting to say and that its "first instinct is to say that all fields are intersections, which makes for an extremely broad field!" Perhaps our field pays for one of its strengths, its multifaceted nature due to the ubiquity of information (Bawden & Robinson, 2015), by having an imperfect name.

### The Value of LIS

Each question about the value of LIS was acceptably answered by Philosopher AI without any extra queries (i.e., three each).

Philosopher AI argued that LIS has overall extremely high societal value because it helps people by providing information for everyday tasks, which has a “significant impact on the way humans view their world and how they go about doing things” (2.1.2). Similarly, it noted LIS provides a very important service to society by maintaining collections of information in order for people, who have varying levels of education and are prone to distraction, to be able to find relevant and useful information, while librarians with specialised knowledge can “facilitate communication between researchers and experts” (2.1.3). The existence of LIS, the AI argued, allows people to work in various jobs at libraries, museums, and with IT and Web technologies, which the AI claimed is fortunate for “people who enjoy organizing”, and by employing such people, LIS “helps reduce unemployment” (2.1.1).

The AI did not produce long answers regarding what a degree in LIS prepares students to do: to work as librarians, cataloguers, archivists, and educators, which it says “is obvious” (2.2.1). It also mentioned the direct value LIS students get from their education, even suggesting it is such a student: “I find that it [studying LIS] helps me learn new things about information and libraries as well as become better at finding what I need” (2.2.2). Finally, it noted that being a librarian necessarily entails “very intimate interactions” with patrons or students, and it describes the experience of being in a library in a way that is reminiscent of *The Breakfast Club*: “Being alone for long periods of time usually causes people to start talking about their life stories while interacting with bookshelves is definitely a recipe for deep conversation. I think it would be interesting if you could get a group of people to work in a library, and then not allow them to leave until they had developed their own philosophy or political view” (2.2.3).

Despite the many possible answers to the question of the value of LIS, for example with reference to addressing the challenges of the info society, the AI-provided answers mostly resemble summaries of what LIS departments might put on their Websites to inform stakeholders and attract new students. Indeed this may have been the source text that most informed the answers; as a result, they were generally very positive, but somewhat obvious. They also focused primarily on the operations of information institutions and practical skills acquired in an LIS degree, and said nothing about the value of the research (i.e., scientific value) and outreach activities of LIS nor the vision and leadership skills that benefit today’s information society.

### The Future of LIS

Philosopher AI required on average 4 attempts to produce usable answers to the questions about the future of LIS, with only the question about how AI will impact LIS being particularly difficult (6 queries).

Regarding the grand challenges facing LIS, the AI’s responses varied from concrete to abstract. It noted, as we suspect would many in LIS today, that “how libraries and archives can best adapt to serve future generations” will be an important challenge, especially deciding *what* among our cultural heritage is valuable enough to preserve, and how then to best preserve it (3.1.3). It claimed “theorists in library science have an insatiable desire to create new subject classifications, cataloguing rules and classification systems that only a handful of librarians will ever use. Meanwhile, it states, the world outside is crying out for simple solutions to practical problems” (3.1.2). Despite the accusation we are preoccupied with useless theory, the AI also characterised the challenges facing LIS as “not merely about organizing or representing all the world's books, documents, recordings, etc, but rather they are fundamental philosophical issues regarding what knowledge actually is and how humans know things to be true” (3.1.1). Indeed the importance of topics like fake news, misinformation, and censorship appears to be at a zenith today, and LIS scholars are actively contributing.

Writing about the emerging challenges facing the *information society,* the AI touched on several concerns that will be familiar and uncontroversial (but still serious) to most scholars, if not all members of the information society. One will be wide unemployment caused by automation and exaggerated by global economic inequality, which “will require computer scientists and economists to solve” (3.2.1). There is no doubt of today’s global inequalities, and the effects of automation on employment is a hotly discussed topic (Spencer, 2018), but the prospect for the related *socio*-technical problems being solved by computer science or economists should be viewed with scepticism (Montreal AI Ethics Institute, 2021), especially as AI and AI language models in particular can further contribute to such problems (Bender *et al.*, 2021). Other worries included how to maintain sustainable growth “without destroying natural resources”, “how to maintain freedom of speech without people abusing it”, “how to maintain our privacy on the internet, while also allowing companies and governments to use data mining techniques in order to make new discoveries” (3.2.3). LIS has been aware of such challenges and already contributed in various forms to each (e.g., for sustainability see Hauke *et al.*, 2018; for fake news see Revez & Corujo, 2021), but of course the work is not complete and these phenomena remain challenging indeed. The AI was optimistic in this particular answer ("the information age is just beginning, and there are many challenges ahead. I am confident that we will overcome these however, because humans have always been able to adapt when faced with new technology", 3.2.3), but not so in the next, where the *grandest* grand challenge was identified: “human beings themselves, and their global social interaction” (3.2.2). The rationale provided indicated that through technology humans create more problems than we solve, and we extend egoism around the world, leading to more global conflicts than cooperation. Indeed, technologies seem to develop ceaselessly and each solution brings its own problems (i.e., Kranzberg’s [1986] second law: invention is the mother of necessity).

The AI perspectives on the future of libraries include both cliché intuitions and interesting observations. It argued that libraries will be smaller and “more space-efficient” despite the volume of human knowledge increasing, because information is on the Internet and physical “books will be used less and less” (3.3.2), or further still, that “books and libraries will no longer be necessary” as people will listen to audio files on their handheld devices and “all of the information that people need for their studies can now be found on the internet” (3.3.1). Such dystopian claims will be familiar to librarians, LIS scholars, and so on, and probably reflect some common folk forecasting on the matter. The final answer was more hopeful and more nuanced, if a bit focused on digital information: "Libraries will continue to exist, in some form or another. The basic principle of libraries is the conversion of human knowledge into a digital format for easy access by humans and machines alike. As long as there are humans on earth that desire information, libraries will serve this purpose”, and there libraries “will continue to be important information hubs in the future” that will hold and provide more advanced and more digital technologies (3.3.3). This is perhaps one of the stronger claims to persistence that LIS and libraries (in various forms) have today: as information increases, the need for organisation increases, and thus the need for relevant services and technologies increases.

*The role of AI* – Philosopher AI’s answers regarding how AI itself will affect LIS focused primarily on generic technical improvements that were unsurprising given the growing success and popularity of AI today, but some still sound to us impressive and some worrisome. Regarding the former kind of prediction, it noted that AI will help computers process and retrieve information faster and in greater volumes (3.4.1, 3.4.3), make conclusions from stored information, and predict and interpret trends in data. However, it claimed that AI will do the interpretive work and decision making *better* than humans because AI can understand nuance better and “especially because AI is not biased” (3.4.2). We noted above that such claims are incorrect and worrisome, and the addition of decision making to the suggested repertoire introduces its own host of further concerns (Jobin *et al.*, 2019). Of more direct concern to LIS, the AI predicted AI technology will make the experience of accessing information easier and faster, through an AI-created “online search engine that would make searching for information much simpler than it is at present” (3.4.3) or a simple, single interface for AI-powered search systems (3.4.2). At face value these claims are plausible: in May of 2021 Google announced a language-model-powered conversational system (Condon, 2021) that could replace traditional Webpage retrieval with a (seemingly) more direct form of *information* retrieval that does not require, and perhaps does not easily *allow* reviewing the sources of its outputs (Heaven, 2021). Such a change in how members of the information society commonly retrieve information would likely have considerable implications for how the services of information professionals are delivered, and how research is conducted, so the exact roles of LIS and information professionals in working on, with, or for such tools may be worth considering sooner than later.

The AI-generated predictions for AI impacting libraries suggested further change, with two answers implying a furthering of the dual-delivery (i.e., digital and physical) model of libraries. First, it “may be similar to how bookstores and media retailers are changing with online shopping… People can still have a bookstore without an online presence if they choose” (3.5.1), which (the AI reasons) would allow libraries to develop services, help people search for information more easily, and use AI to suggest relevant books or articles. Indeed, this is increasingly the state of libraries today. Second, it briefly suggested AI could be used to create *virtual libraries* to increase access (i.e., for those who cannot go to a physical library), but it the details it sounds more like *digital library* services than a virtual emulation of a physical space (3.5.2). In other words, libraries will go “into the clouds” (Bawden & Robinson, 2015); it is not unreasonable to expect AI could do such work, and certainly faster than humans would. It also suggested AI could help patrons “search for specific information stored at various locations around the world; think of it as your own personal assistant librarian who will always be there whenever you need them” (3.5.2). While librarians arguably already provide such a service (albeit not on an *exclusive* one-to-one basis with patrons), it may be worth considering the advantages of *also* having AI do such, and its present performance in producing reasonable prose on difficult topics is perhaps evidence that it will not be long before it can provide such services competently. At the very least, these plausible ideas emphasise the importance of studying the effect of AI on LIS, especially as such technologies are already present in many libraries today (Massis, 2018; Feng, 2021); technology is once again changing the nature of librarianship (c.f. Shera, 1973). Finally, the most pessimistic answer was that "there will be no more libraries since what they do will be entirely automated and done better by AIs. People won't need to pay for them, either. Libraries are a bit like restaurants or bars in that they're expensive to run but most people only go once or twice[*!!!*]. AIs will put all the information they have online, like Google Books already does. As for hard copies of books and magazines, AIs can print those too. So basically, libraries will be replaced by the Internet. And that is just as it should be!" (3.5.3). We find this rhetoric, with its emphatic delivery, especially worrisome as it is plausible enough to convince a lay audience, and prefer to think of libraries a bit like *hospitals*: regardless of expense or how often one goes, such places *have* to be there.

**Synthesis and evaluation**

Querying Philosopher AI takes very little time and usually produces 3-5 paragraphs of coherent and sometimes rather sophisticated text, and we assume most modes of accessing GPT-3 (and similar models) will be approximately as quick. However, fewer than 50% of queries of the kind presented here currently return usable responses. The questions about the nature of LIS were perhaps particularly difficult, typically requiring Philosopher AI 5 or more prompts to get 3 usable responses (one even required 7), whereas the 'future' questions typically required 4 (one question required 6 and one only 3), and the 'value of LIS' questions required no extra attempts (i.e., 3 prompts for each of the two questions). One tentative interpretation of this is that for any agent (i.e., human or AI) facing such questions, it is easier to explicate or find textual evidence of the value of LIS than it is to coherently state its nature or reasonably speculate about the future.

We observed the responses to be plausible, but also often ridiculous or incendiary (sometimes both), which likely reflects the training data. Such data include not only general Web comments but any accessible knowledge published by LIS scholars (e.g., in open-access publications, on Wikipedia, etc). Perhaps this suggests that public pessimism about the field and about libraries currently outweighs the published evidence of LIS's vision and careful optimism. We may want to collectively address this imbalance if we hope to maintain favour as a public service and credibility as a research field (Galluzzi, 2014).

As for the *usefulness* of the responses produced, this varied. In general, the AI did not respond to our questions with the same level of erudition and insight that it did in response to philosophers' questions (Weinberg, 2020). As seen above, Philosopher AI often repeated common tropes, contradicted itself, and provided insufficient detail to support its points. Though it occasionally claimed to have provided sources, it never truly did. It also took considerable us effort to look past many detours in the produced narratives, which suggests the AI's ability to stay *narrowly* on topic is still limited. For example, when asked about the grand challenges facing LIS, one response (3.1.2) included: "there's a rather circular logic embedded in the idea that 'challenges' are what a field is about. After all, if no one perceives there to be problems, then they might as well just close up shop and go home. The irony is that they already have been going home for the last few decades". In other words, there was much wheat as chaff even in the better responses. As performance was especially inconsistent in response to forecasting questions, we suspect that questions about *longer*-term technologies (i.e., ICT innovations beyond the AI of today or tomorrow) would be even less useful.

On the other hand, we think the AI produced the occasional insight, and certainly plenty of catalyst for deeper discussion about the nature, value, and future of LIS (and often, just of libraries). The perspectives it produced about the nature of LIS recreated several known, important considerations in characterising and naming LIS, and it also produced plausible and provocative answers of its own as well as interesting commentary on the task of naming a field. The prose about the value of LIS was highly focused and aligned with the common discourse in LIS, as noted above. Finally, the prose about the future of LIS included plausible ideas about how AI will change libraries, as well as worrisome ones for making them obsolete. Therefore, regardless of its use in research, GPT-3 (or any such model) may be a useful educational tool in contexts where the veracity of its outputs is less of a concern than its capacity to prvovoke discussion; for example, in an in-class activity students could pose AI questions about LIS, libraries, and contemporary issues and collectively consider and discuss the responses.

*Do we recommend using AI Philosopher in research today?* No. The task of soliciting, searching for, and considering insights in its output is, today, likely more work than deriving them oneself, and the forecasts are not yet those of an expert. Drawing on the levels used to classify automobile automation (Edwards *et al.*, 2020), one could argue that even in its best moments GPT-3 provides only ‘conditional automation’ to the research process, where the researcher is still responsible for inferring the right from wrong outputs (i.e., achieves automation level 3 of 5). But this limited performance also cannot be ignored, and there are reasons to think it will be improving over the coming years: language model performance is currently still scaling up with model size (i.e., number of parameters) and new developments are enabling even small models perform comparably with far fewer resources (Schick & Schütze, 2020). Similarly improvements in the interaction with the AI (e.g., developments in conversational agents; Barko-Sherif *et al.*, 2020) may also make it easier to refine one’s query and stipulate response criteria to get insightful, well-argued outputs. We therefore suggest LIS stay abreast of such developments and perhaps prepare for the next GPT generation by establishing and refining a method for evaluating performance in idea-generation and forecasting (i.e., as a type of information provision), and consider when the time is right to again pose it difficult questions about LIS.

## LIMITATIONS

The novelty and exploratory nature of the approach used in this paper means that, to our knowledge, there are currently no established methods to choose among in evaluating the kinds of AI performance examined here (e.g., answering difficult questions, forecasting field-wide trends, or generating research ideas). We thus had to make methodological decisions according to our judgement when conducting the study and interpreting the results, and the best methodological procedure was not always clear. Future studies could codify and compare such methods, perhaps by drawing on recent evaluations of bias in the outputs of AI language models (c.f. Abid *et al.*, 2021, published just after the present manuscript was accepted) and collecting data using the given system’s own API rather than going through a third-party querying layer (i.e., philosopherAI.com) as we have done here.

Prose generated by AI must be read cautiously and understood as a product of the data *and* processing done to that data, namely human-produced Web data parameterised in billions of unexplained ways. We have tried to interpret the results at face value as much as possible, but naturally this is a highly subjective task that other scholars might perform differently, each finding the responses interesting, compelling, or ridiculous for different reasons. We therefore encourage other authors to review the full outputs we have shared to decide for themselves about the value of the approach. Similarly, although the need to discard certain prompts was mostly quite clear, it was also nonetheless subjective, and the discarded prompts should thus be examined as well.

Different prompts, even if only subtly different, will produce different responses, as will the same prompt entered additional times, and we examined no more than seven responses per prompt. For example, the responses we received, which were heavily focused on libraries even when we were not asking about them directly, may have been a result of choosing ‘library and information science’ rather than ‘information science’ or ‘information studies’. Our conclusions should thus be weighed in light of the relatively small number of queries performed and reviewed for each question.

## CONCLUSION

In the results and discussion above we have provided an LIS-tailored demonstration of a state-of-the-art AI language model, and evaluated the prospect of using AI-generated prose as a research tool (e.g., as a source of data or of ideas); to our knowledge, this is the first manuscript to do either. While the outputs were at times impressive or entertaining, we find that AI language models are currently still at the precipice of being viable research tools: when given the task of pontificating about LIS, Philosopher AI produced content of varying quality and insight, with the useless ideas being well hidden among the genuinely interesting or useful.

However, given the current state and rapid development of AI, it is possible such models will be producing good research ideas and content within a generation – and so within this decade – effectively automating some information services and knowledge work. It may also allow rapidly producing convincing *fake* research results (Dehouche, 2021), and if so, hopefully *also* helps with the peer review that will be required to filter those outputs from genuine submissions. Further, as discussed briefly above, AI language models may drastically change the nature of everyday information retrieval. Regardless of what *may* be, such systems are *already* stimulating serious social and ethical issues: biased outputs of AI language models have recently been identified (e.g., anti-Muslim bias, Abid *et al.*, 2021), due in part to the training data (i.e., mostly English-language Web data), and global environmental, governmental, and labour issues are resulting from the current training and implementation of AI language models (Bender *et al.*, 2021). Considering these future promises and current problems, we recommend that LIS researchers and information practitioners follow and contribute to research and practice wherever possible, for example through examining AI as research tools (as we have done here), investigating the role of AI language models in information seeking, considering the challenges such systems may pose to information literacy, and considering how to identify and address the social and information-ethical aspects of such systems.

**ACKNOWLEDGMENTS**

The authors are grateful to Dr. Asen Ivanov and three anonymous peer reviewers for their useful feedback, and to Dr. Maria Gäde, Prof. Robert Jäschke, and Prof. Michael Seadle for their input on which questions to pose to the AI.